\documentclass{elsart}

\usepackage{graphicx}
\usepackage[dvips]{thumbpdf}
\usepackage{amsmath,amssymb,amsfonts}
\usepackage{cite}

\newcommand{\ssst}{\scriptscriptstyle}
\newcommand{\sst}{\scriptstyle}
\newcommand{\dst}{\displaystyle}

\begin{document}

\begin{frontmatter}

\title{Electroproduction of kaons from the proton \protect\\ in a Regge-plus-resonance approach}

\author{T. Corthals}, 
\author{T. Van Cauteren}, 
\author{P. Vancraeyveld}, 
\author{J. Ryckebusch}\ead{jan.ryckebusch@ugent.be}
\address{Department of Subatomic and Radiation Physics, 
Ghent University, Proeftuinstraat 86, B-9000 Gent, Belgium}
\author{D.G. Ireland}
\address{University of Glasgow, Glasgow G12 8QQ, United Kingdom}

\begin{abstract}
We present a Regge-plus-resonance (RPR) description of the $p(\mathrm{e},\mathrm{e}'K^+)Y$ processes ($Y=\Lambda,\Sigma^0$) in the resonance region. The background contributions to the RPR amplitude are constrained by the high-energy $p(\gamma, K^+)Y$ data. As a result, the number of free model parameters in the resonance region is considerably reduced compared to typical effective-Lagrangian approaches. We compare a selection of RPR model variants, originally constructed to describe $KY$ photoproduction, with the world electroproduction database. The electromagnetic form factors of the intermediate $N^{\ast}$s and $\Delta^{\ast}$s are computed in the Bonn constituent-quark model. With this input, we find a reasonable description of the $p(\mathrm{e},\mathrm{e}'K^+)Y$ data without adding or readjusting any parameters. It is demonstrated that the electroproduction response functions are extremely useful for fine-tuning both the background and resonant contributions to the reaction dynamics. 
\end{abstract}

\end{frontmatter}

The electromagnetic production of mesons from the nucleon is widely envisaged as a stepping-stone to linking partonic and hadronic degrees of freedom. Study of the strange $KY$ channels, in particular, is expected to yield insight into issues such as the flavor dependence of the strong interaction and the search for missing resonances~\cite{CapRob94-Loe01}. A thorough grasp of the $p(\gamma^{(\ast)},K)Y$ reaction dynamics is also essential for hypernuclear production, a field which has been rapidly gaining momentum over the past few years. 

Recent measurements at the JLab, ELSA, SPring-8 and GRAAL facilities have resulted in an impressive set of high-precision $p(\gamma^{(\ast)},K)Y$ data in the few-GeV regime~\cite{McNabb04,Bradford06,Glander04,Lawall05,Zegers03,Sumihama05,Graal06,Schum06,Mohring03,Carman03,Carman06}. This has triggered renewed efforts by various theoretical groups, including the develop\-ment of tree-level isobar models, in which the amplitude is constructed from lowest-order Feynman diagrams~\cite{AdWr88,WJ92,DaSa96,MaBe00,Stijnprc01,Sara05}, and of more elaborate coupled-channels approaches~\cite{US05,MoShklyar05,Diaz05}. 
While most of these analyses center on the real-photon process, which heavily dominates the current dataset, it has been shown that the electroproduction observables can yield important complementary insights~\cite{Stijn_el}. As a combined coupled-channels analysis of the $p(\gamma,K)Y$ and $p(e,e'K)Y$ reactions has not yet been implemented, a tree-level approach currently represents the best possibility of studying both reactions within the same framework.

A major challenge for any description of electromagnetic $KY$ production is parameterizing the nonresonant contributions to the amplitudes. Over the years, several background models have been suggested~\cite{WJ92,MaBeHy95,Stijnprc01}, differing primarily in the mechanism used to reduce the Born-term contribution, which by itself spectacularly overshoots the measured cross sections. While all models are able to provide a fair description of the data, it turns out that the extracted resonance couplings depend strongly on this choice~\cite{GApaper03}.

In Refs.~\cite{CorthalsL,CorthalsS}, we have developed a tree-level effective-field model for $\Lambda$ and $\Sigma$ photoproduction from the proton. It differs from traditional isobar approaches in its description of the nonresonant diagrams, which involve the exchange of kaonic Regge trajectories in the $t$ channel~\cite{reg_guidal}. This Regge background is supplemented with a selection of $s$-channel resonances. Such a ``Regge-plus-resonance'' (RPR) strategy has the advantage that the background diagrams contain only a few parameters, which can be constrained by the high-energy data. Further, the use of Regge propagators eliminates the need to introduce strong form factors in the background terms, thus avoiding the gauge-invariance issues plaguing traditional effective-Lagrangian models~\cite{US05}. In this Letter, we use the RPR model variants constructed in Refs.~\cite{CorthalsL,CorthalsS} to obtain predictions for the $p(\mathrm{e},\mathrm{e}'K^+)\Lambda,\Sigma^0$ processes. It will be shown that the electroproduction response functions are very useful for fine-tuning certain model parameters which the photoproduction data fail to constrain. We focus our comparisons on the newly released CLAS data from Ref.~\cite{Carman06}. \\

In a Regge model, the reaction dynamics are governed by the exchange of entire Regge trajectories rather than of single particles. An efficient strategy is to embed the Regge formalism into a tree-level effective-field model, as proposed in~\cite{reg_guidal,Guidal_elec}. Here, we consider the exchange of kaonic trajectories in the $t$ channel. The amplitude for exchange of a linear kaon trajectory $\alpha_X(t)=\alpha_{X,0} + \alpha'_X \, (t-m_X^2)$, with $m_X$ and $\alpha_{X,0}$ the mass and spin of the trajectory's lightest member $X$, can be obtained from the Feynman amplitude by replacing the Feynman propagator with a Regge one:
\begin{equation}
\frac{1}{t-m_X^2} \hspace{6pt} \rightarrow{~\quad} \hspace{8pt} \mathcal{P}^X_{Regge}[s,\alpha_X(t)]\,.
\end{equation}
The Regge amplitude can then be written as
\begin{equation}
\mathcal{M}^X_{Regge}(s,t,Q^2) = \mathcal{P}^X_{Regge}[s,\alpha_X(t)] ~\times~ \beta_X(s,t,Q^2) \,,
\label{eq: define_reggeprop}
\end{equation}
with $\beta_X(s,t,Q^2)$ the residue of the original Feynman amplitude. 

In our treatment of $\Lambda$ and $\Sigma^0$ photoproduction \cite{CorthalsL, CorthalsS}, we identified the $K^+(494)$ and $K^{\ast +}(892)$ trajectories as the dominant contributions to the high-energy amplitudes. The corresponding propagators are given by:
\begin{align} 
\mathcal{P}^{K^+(494)}_{Regge}(s,t) = \left(\frac{\dst s}{\dst s_0}\right)^{\alpha_K(t)}
\frac{1}{\sin\bigl(\pi\alpha_K(t)\bigr)} \; \frac{\pi \alpha'_K}{\Gamma\bigl(1+\alpha_K(t)\bigr)} \ \left\{ \begin{array}{c}
1 \\ e^{-i\pi\alpha_{K}(t)} 
\end{array}\right\} \,, 
\label{eq: reggeprop_K}
\end{align}
\begin{align}
\mathcal{P}^{K^{\ast +}(892)}_{Regge}(s,t) = \left(\frac{\dst s}{\dst s_0}\right)^{\alpha_{K^{\ast}}(t)-1} 
\frac{1}{\sin\bigl(\pi\alpha_{K^{\ast}}(t)\bigr)} \; \frac{\pi \alpha'_{K^{\ast}}}{\Gamma\bigl(\alpha_{K^{\ast}}(t)\bigr)} \ \left\{ \begin{array}{c}
1 \\
e^{-i\pi\alpha_{K^{\ast}}(t)}
\end{array}\right\}\,,\label{eq: reggeprop_Kstar}
\end{align}
with $\alpha_{K}(t) = 0.70 \ (t-m_{K}^2)$ and $\alpha_{K^{\ast}}(t) = 1 + 0.85 \ (t-m_{K^{\ast}}^2)$~\cite{CorthalsL}. Either propagator can be used with a constant~(1) or rotating~($e^{-i\pi\alpha(t)}$) phase. 
In addition, in Ref.~\cite{reg_guidal} it is argued that, to impose current conservation, the Regge amplitude should contain the electric contribution to the $s$-channel Born term~(i.e.~the part $\sim~e \overline{N} \gamma_{\mu} N A^{\mu}$~\cite{CorthalsL}), leading to:
\begin{align}
&\mathcal{M}_{Regge}\,(\gamma^{(\ast)}\,p \rightarrow K^+ \Lambda,\Sigma^0) = \mathcal{M}_{Regge}^{K^+(494)} + \nonumber \\
&\ \mathcal{M}_{Regge}^{K^{\ast +}(892)} + \mathcal{M}_{Feyn}^{p\ssst,\sst elec} \times \mathcal{P}_{Regge}^{K^+(494)} \times (t-m_{K^+}^2).\label{eq: gauge_recipe}
\end{align}
The unknown coupling constants and trajectory phases contained in $\mathcal{M}_{Regge}$ can be determined from the high-energy $p(\gamma,K^+)Y$ data~\cite{CorthalsL,CorthalsS}. 

Eq.~(\ref{eq: gauge_recipe}) applies to photo- and electroproduction in the  high-energy region. At lower energies, the observables exhibit structures which can be described by supplementing the reggeized background with a number of resonant $s$-channel diagrams. For the latter we assume standard Feynman propagators, in which the resonances' finite lifetimes are taken into account through the substitution $\,s - m_{R}^2 \rightarrow ~ s - m_{R}^2 + im_{R}\,\Gamma_R\,$ in the propagator denominators, with $m_R$ and $\Gamma_R$ the mass and width of the propagating state ($R=N^{\ast},\Delta^{\ast}$). It is required that the resonant diagrams vanish at large values of $\omega_{lab}$. This is accomplished by including a Gaussian hadronic form factor at the $KYR$ vertices: $F(s) = \exp\,[- (s-m^2_{R})^2/\Lambda_{res}^4]$. Our motivation for assuming a Gaussian shape is explained in Ref.~\cite{CorthalsL}. A single cutoff mass $\Lambda_{res}$ is assumed for all $N^{\ast}$s and $\Delta^{\ast}$s. Along with the resonance couplings, $\Lambda_{res}$ is used as a free parameter when optimizing the model against the resonance-region data. 

The relevant strong and electromagnetic interaction Lagrangians are contained in Ref.~\cite{CorthalsL} for the photoproduction case. We assume that the electroinduced processes can be described by the same type of reaction amplitudes, modified with suitable electromagnetic form factors (EMFFs). 

For the $K^+(494)$ and $K^{\ast +}(892)$ trajectories, a monopole EMFF $F_{K^+,K^{\ast +}}(Q^2) = (1 + Q^2/\Lambda_{K^+,K^{\ast +}}^2)^{-1}$ is assumed, with $\Lambda_{K^+} = \Lambda_{K^{\ast +}} = 1300$ MeV, in accordance with Ref.~\cite{Guidal_elec}. The cutoff values were chosen to optimally match the behavior of the electroproduction data in the high-$Q^2$ ($Q^2 \gtrsim 2.5$ GeV$^2$) region~\cite{Mohring03}, where resonant contributions are small. Since the $s$-channel term of Eq.~(\ref{eq: gauge_recipe}) is essentially an artefact of the gauge-breaking nature of the $K^+$-exchange diagram, the most natural way to guarantee current conservation for $Q^2\neq 0$ is to adopt the same EMFFs at the $\gamma^{\ast} p p$ and $\gamma^{\ast} K^+ K^+$ vertices. As pointed out in Ref.~\cite{reg_guidal}, this is also a necessary condition for reproducing the measured $\sigma_L/\sigma_T$ ratios. 

Instead of employing the standard phenomenological dipole parameterizations for the $N^{\ast}$ and $\Delta^{\ast}$ EMFFs, we use those obtained within the covariant constituent-quark model (CQM) developed by the Bonn group~\cite{bonn}. The seven parameters of this CQM have been fitted to the baryon spectrum. No new parameters are introduced when computing the EMFFs. The CQM results compare favourably to the existing data on helicity amplitudes for the low-lying $N^{\ast}$ and $\Delta^{\ast}$ states. However, very few data are available for resonances in the mass region of interest to kaon production ($m_R \gtrsim 1.6$~GeV). We deem that by using computed EMFFs instead of dipoles, we reduce the degree of arbitrariness in the $p(\mathrm{e},\mathrm{e}'K^+)Y$ framework. 

In Refs.~\cite{CorthalsL,CorthalsS}, we constructed RPR amplitudes for the various $\gamma p \rightarrow KY$ channels. A number of variants of the RPR model were found to provide a comparably good description of the $\Lambda$, $\Sigma^0$ and $\Sigma^+$ photoproduction observables. Their properties are listed in Table~\ref{tab: kl_models}. All models include the known $S_{11}(1650)$,  $P_{11}(1710)$, $P_{13}(1720)$ and $P_{13}(1900)$ resonances. Apart from these, each $K^+\Lambda$ variant assumes either a missing $D_{13}(1900)$ or $P_{11}(1900)$, following suggestions from Refs.~\cite{MaBe00,GApaper04,MaSu06}. The $K^+\Sigma^0$ amplitude further contains the $D_{33}(1700)$, $S_{31}(1900)$, $P_{31}(1910)$ and $P_{33}(1920)$ $\Delta^{\ast}$ states. A good description of this channel could be achieved without the introduction of any missing resonances. The parameters of the $K^+\Lambda$ variants from Ref.~\cite{CorthalsL} have been readjusted in order to reproduce the recent beam- and recoil-asymmetry data from GRAAL~\cite{Graal06}.

The background contribution to the RPR amplitude involves three parameters: one for the $K(494)$ trajectory ($g_{KYp}$) and two for the $K^{\ast}(892)$ trajectory ($G^v_{K^{\ast}}$ and $G^t_{K^{\ast}}$, corresponding to the vector and tensor couplings~\cite{CorthalsL}).  Their values were determined through a fit to the high-energy ($\omega_{lab} \gtrsim 4$ GeV) observables. As can be appreciated from Table~\ref{tab: kl_models}, it turned out to be impossible to fix the sign of $G^t_{K^{\ast}}$ with the available photoproduction data. Furthermore, for each trajectory propagator, either a constant (cst.) or rotating (rot.) phase may be assumed. In the $K^+\Lambda$ channel, two combinations (rot. $K$, rot. $K^{\ast}$ and rot. $K$, const. $K^{\ast}$) produce a comparable quality of agreement between the calculations and the combined high-energy and resonance-region data. With respect to the quantum numbers of a potential ``missing'' $N^{\ast}(1900)$ resonance, both $P_{11}$ and $D_{13}$ emerged as valid candidates. 


\sloppy 

One issue that may cloud the proposed RPR strategy is double counting, which could result from superimposing a number of individual resonances onto the reggeized high-energy background. It has been found that hadronic scattering amplitudes exhibit the property of duality~\cite{Dolen68}, as quantified by finite-energy sum rules~\cite{Collins77}. The latter can be understood to signify that the sum of all resonances in the $s$ channel, when averaged over energy, equals the sum of all Regge-trajectory exchanges in the $t$ channel. It remains unclear, however, how this ``reggeon-resonance'' duality can be implemented into a meson-photoproduction reaction model.\footnote{For meson-meson scattering, on the other hand, the pioneering approach of Veneziano~\cite{Ven68} and related ``dual-resonance'' models have proven quite successful~\cite{Man92}.}

\fussy 


In what follows, we will present a method to estimate the effect of double counting in our RPR model. 
Duality implies that the ($t$-channel) background couplings, which have been constrained against the high-energy data, may carry certain resonant contributions. 
To remedy this, one may opt to re-fit the background parameters when addressing the resonance-region data. 
The essential question, then, is how strongly the resonance and background couplings are affected by such a procedure.  

\begin{figure*}
\begin{center}
\includegraphics*[width=0.58\textwidth]{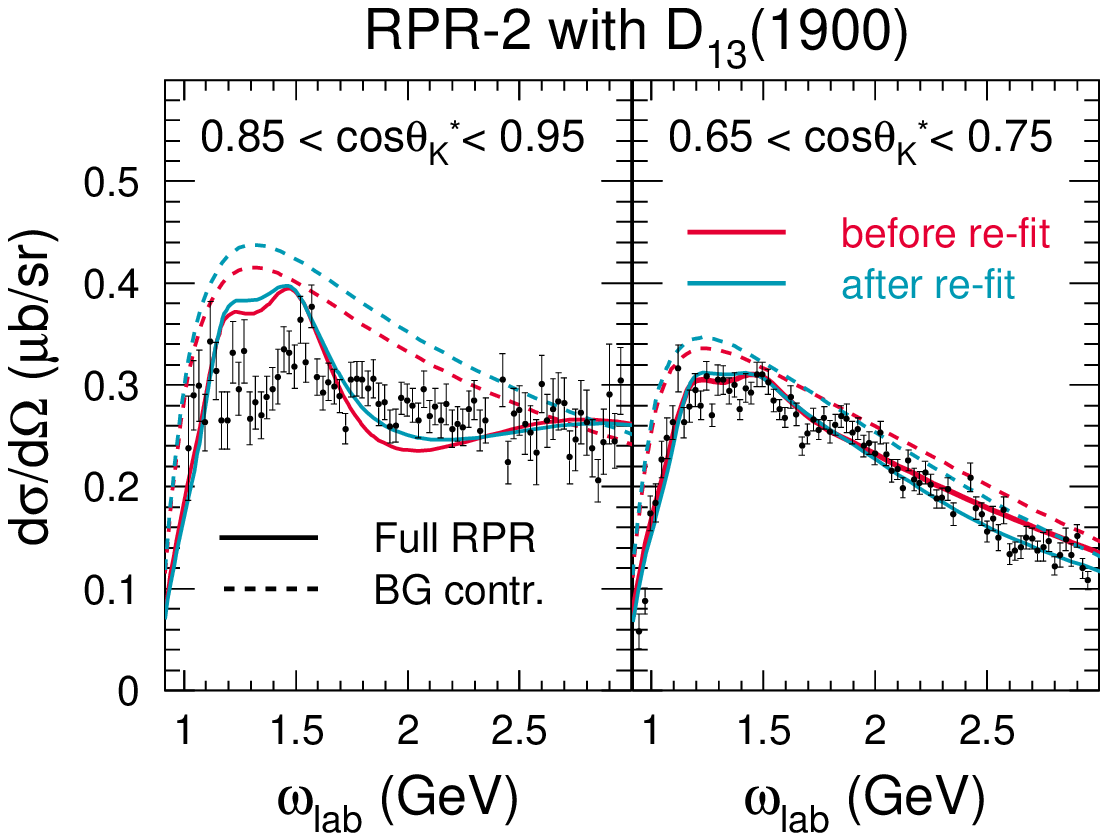}
\end{center}
\caption{Forward-angle $p(\gamma,K^+)\Lambda$ differential cross section for the RPR-2 model with a missing $D_{13}$, before (red curves) and after (blue curves) re-fitting the background and resonance couplings to the resonance-region data. The full and dashed curves correspond to the full RPR amplitude and its background (BG) contribution, respectively. The values of $\chi^2_{\mathrm{RPR}}$ before and after re-fitting are 2.7 and 2.4.}
\label{fig: pho_comp_doubcount}
\end{figure*}

Figure~\ref{fig: pho_comp_doubcount} compares a selection of $p(\gamma,K^+)\Lambda$ differential cross sections for 
the RPR-2 model with a missing $D_{13}$, before and after re-fitting the background (\emph{and} resonance) couplings to the resonance-region data. 
It is clear that neither the background contribution nor the full RPR result are considerably affected by this re-fitting. The impact on the extracted values of the model parameters is, however, non-negligible. 
Specifically, we observe variations in the fitted coupling constants which range between 5 and 35\%. The extracted background parameters turn out to be slightly more stable than the resonant ones. 

From the above analysis, we conclude that the running of the background couplings with the energy scale is relatively modest. Further, we estimate that double-counting effects give rise to 20\% errors on the extracted resonance parameters. It is worth remarking that Chiang \emph{et al.} reached a comparable conclusion in their RPR description of $\eta$ and $\eta'$ photoproduction~\cite{Chiang}. 

\begin{figure*}
\begin{center}
\includegraphics*[width=0.63\textwidth]{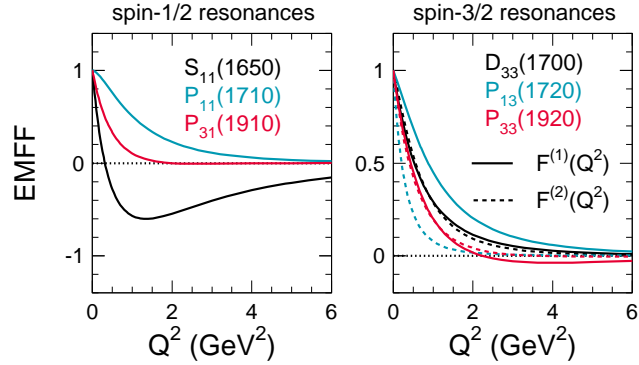}
\end{center}
\caption{EMFFs for the $\gamma^{\ast} p R$ ($R=N^{\ast},\Delta^{\ast}$) interactions as computed in the Bonn CQM. For spin-1/2 states, the EM vertex only contains a single term, whereas for spin-3/2 states, $F^(1)$ and $F^(2)$ correspond to the first and second terms in the EM Lagrangian from Ref.~\cite{CorthalsL,CorthalsS}.}
\label{fig: bonn_emffs}
\end{figure*}

After multiplying the EM couplings with the necessary EMFFs, we compare the RPR variants from Table~\ref{tab: kl_models} with the electroproduction observables. When assuming the effective Lagrangians from Refs.~\cite{CorthalsL,CorthalsS}, the spin-1/2 and spin-3/2 resonances acquire one and two EMFFs, respectively. Figure~\ref{fig: bonn_emffs}  displays the Bonn CQM results for the $S_{11}(1650)$, $P_{11}(1710)$, $P_{13}(1720)$, $D_{33}(1700)$, $P_{31}(1910)$ and $P_{33}(1920)$ EMFFs. As the computed form factors of the $P_{13}(1900)$ turned out to be too small, we used a standard dipole shape with a cutoff of $840$~MeV~\cite{Stijn_el}. The same parameterization was adopted for the $S_{31}(1900)$, the mass of which is overestimated in all existing CQMs, and for the missing $D_{13}(1900)$ and $P_{11}(1900)$. 

We use the following definition for the unpolarized responses:
\begin{equation}
\frac{\mathrm{d}\sigma}{\mathrm{d}Q^2 \mathrm{d}W \mathrm{d}\Omega^{\ast}_K} = \Gamma_v \left[ \sigma_T + \epsilon \, \sigma_L + \epsilon \, \sigma_{TT} \cos 2\varphi_K + \sqrt{\epsilon(1+\epsilon)} \, \sigma_{LT} \cos\varphi_K \right],
\end{equation}
with $\epsilon = (1+\frac{2\vert\overrightarrow{q}\vert ^2}{Q^2}\,\tan^2\frac{\theta_e}{2})^{-1}$ and $\Gamma_v$ a kinematical factor, defined for example in Ref.~\cite{Carman06}. For the transferred polarization observables ($P'_{x}$, $ P'_{x'}$, $P'_{z}$ and $P'_{z'}$) the conventions from Ref. \cite{Carman03} are assumed.

While a limited selection of results was already shown in Ref.~\cite{CorthalsHypproc}, the release of new data by the CLAS collaboration~\cite{Carman06} is a unique opportunity to subject the predictive power of the RPR model to a stringent test. Remarkably, the data from Ref.~\cite{Carman06} appear to favor a reggeized description of the $p(\mathrm{e},\mathrm{e}'K^+)Y$ processes. Specifically, the Regge model of Guidal~\cite{Guidal_elec} is found to reproduce the CLAS data consistently better than the isobar models of both Janssen~\cite{Stijn_el} and Mart~\cite{MaBe00}. Although the reasonable performance of the pure Regge description for most observables suggests a $t$-channel dominated process, there are obvious discrepancies between the Regge predictions and the data, indicative of $s$-channel dynamics. The RPR strategy represents an ideal framework to parameterize these contributions. 

\begin{figure*}
\begin{center}
\includegraphics*[width=0.8\textwidth]{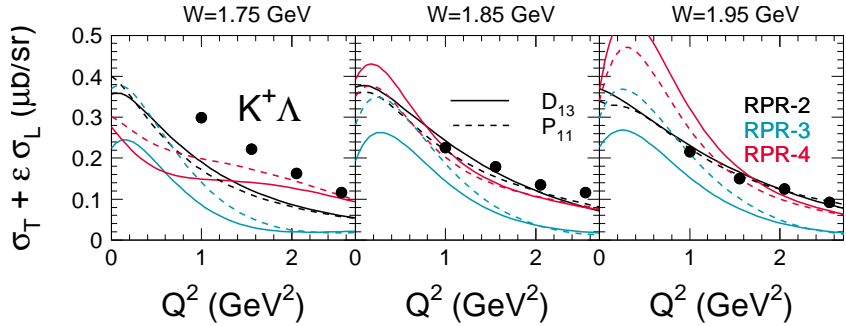}
\end{center}
\caption{$Q^2$ evolution of the unseparated differential cross section $\sigma_T + \epsilon\, \sigma_L$ for the $K^+\Lambda$ final state at $\cos\theta_K^{\ast} = 0.9$, using the model variants from Table~\ref{tab: kl_models}.  The data are from~\cite{Carman06}.}
\label{fig: k+_lambda_su}
\end{figure*}

Figure~\ref{fig: k+_lambda_su} shows the $Q^2$ evolution of the unseparated $p(e,e'K^+)\Lambda$ differential cross sections $\sigma_T + \epsilon\, \sigma_L$, computed with the model variants from Table~\ref{tab: kl_models}. The RPR-3 variant underestimates the data by several factors, whereas RPR-2 and RPR-4 lead to acceptable results at all but the lowest energy. 

\begin{figure*}
\begin{center}
\includegraphics*[width=0.60\textwidth]{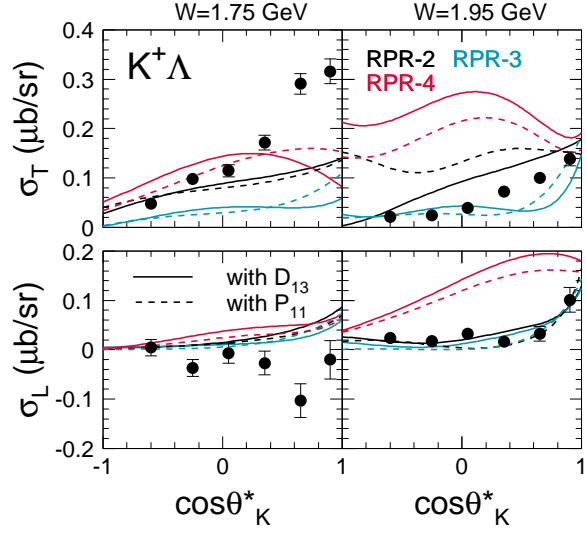}
\end{center}
\caption{$\cos\theta_K^{\ast}$ evolution of the separated differential cross sections $\sigma_L$ and $\sigma_T$ for the $K^+\Lambda$ final state at $Q^2 = 1.0~\mathrm{GeV}^2$, using the six RPR model variants from Table~\ref{tab: kl_models}.  The data are from~\cite{Carman06}.}
\label{fig: k+_lambda_st-l}
\end{figure*}

The separated observables $\sigma_L$ and $\sigma_T$ are shown in Fig.~\ref{fig: k+_lambda_st-l} as a function of $\cos\theta^{\ast}_K$. The longitudinal cross section is clearly the least sensitive to the specific structure of the amplitude, with only the RPR-4 variants failing to reproduce its behavior at higher energies. The transverse cross section is more difficult to describe, as none of the six model variants are able to reproduce its magnitude at forward angles and $W \approx 1.75$ GeV. At higher energies, the RPR-2 variant with a missing $D_{13}(1900)$ performs reasonably, as do both RPR-3 models. The latter two were, however, excluded by comparison with the unseparated data (Fig.~\ref{fig: k+_lambda_su}). 

\begin{figure*}
\begin{center}
\includegraphics*[width=0.85\textwidth]{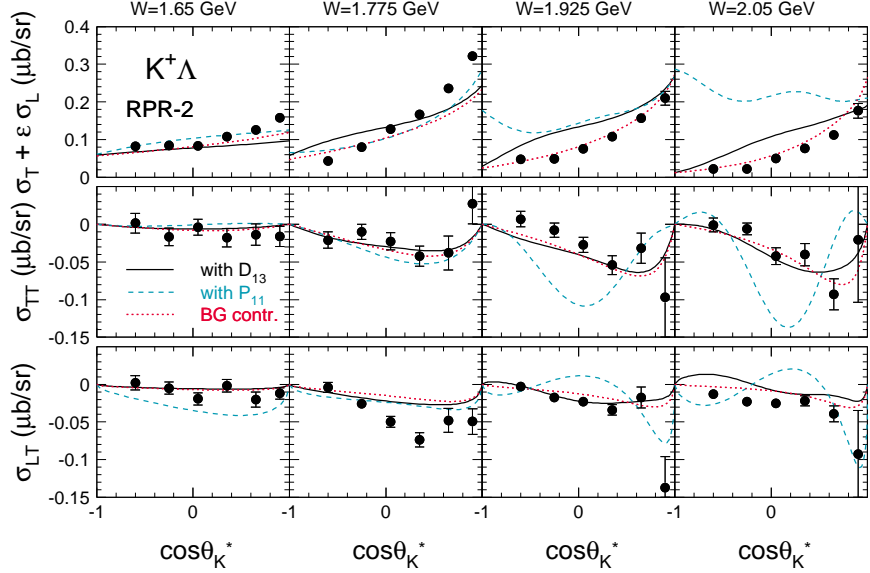}
\end{center}
\caption{$\cos\theta_K^{\ast}$ evolution of the differential cross sections $\sigma_T + \epsilon\, \sigma_L$, $\sigma_{TT}$ and $\sigma_{LT}$ for the $K^+\Lambda$ final state at $Q^2 = 0.65$ GeV$^2$, using the two RPR-2 model variants from Table~\ref{tab: kl_models}.   The dotted curves indicate the contribution of the Regge background, whereas the full and dashed curves correspond to the full amplitudes including a missing $D_{13}$ and $P_{11}$, respectively. The data are from~\cite{Carman06}.}
\label{fig: k+_lambda_s-new}
\end{figure*}

Figure~\ref{fig: k+_lambda_s-new}, which shows the $\cos\theta^{\ast}_K$ dependence of $\sigma_T + \epsilon\,\sigma_L$, as well as of the previously unmeasured observables $\sigma_{TT}$ and $\sigma_{TL}$, supports the above conclusions. 
The RPR-2 variant with a missing $D_{13}$ state reasonably reproduces the trends of the data, including the strong forward-peaking behavior of the unseparated cross section. The variant with a missing $P_{11}$, on the other hand, leads to very poor results for $\sigma_T + \epsilon\,\sigma_L$ and $\sigma_{TT}$. The RPR-4 results (not shown) were also found to deviate strongly from the data, as was the case for Fig.~\ref{fig: k+_lambda_st-l}.

\begin{figure*}
\begin{center}
\includegraphics*[width=0.7\textwidth]{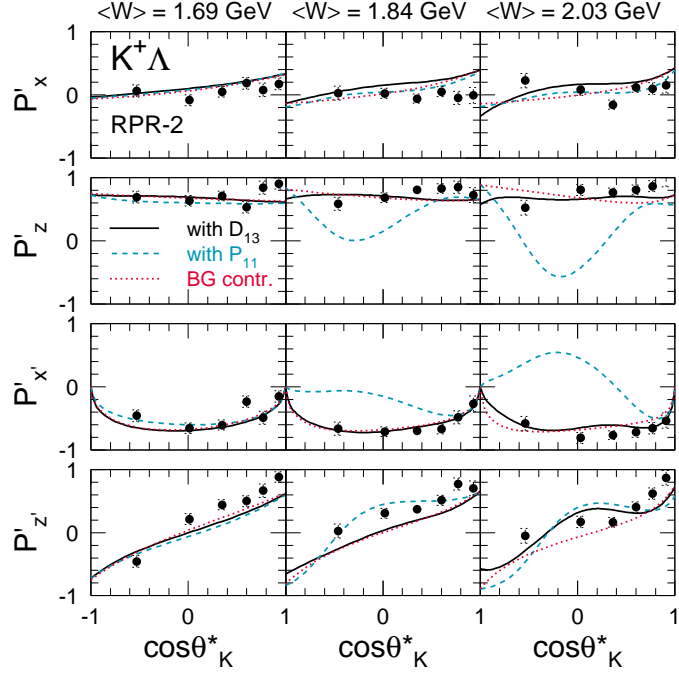}
\end{center}
\caption{Transferred polarization components $P'_x$, $P'_z$, $P'_{x'}$ and $P'_{z'}$ for the $K\Lambda$ final state as a function of $\cos\theta_K^{\ast}$, computed with the RPR-2 model variants from Table~\ref{tab: kl_models}. Line conventions are as in Fig.~\ref{fig: k+_lambda_s-new}. The data are from~\cite{Carman03}.}
\label{fig: k+_lambda_pol}
\end{figure*}

Figure~\ref{fig: k+_lambda_pol} compares our results for the transferred polarization observables $P'_x$, $P'_z$, $P'_{x'}$ and $P'_{z'}$, obtained with either RPR-2 variant, to the data of Ref.~\cite{Carman03}. Once more, it turns out that the $P_{11}(1900)$ option can be discarded. The RPR-2 variant with a missing $D_{13}$ again provides a fair description of the data, and clearly represents the optimum choice for describing the combined photo- and electroproduction processes. This result supports the recent conclusion from Ref.~\cite{MaSu06} that a $D_{13}$ state with a mass around 1920 MeV is required by both the CLAS and SAPHIR $p(\gamma,K^+)\Lambda$ data.

It is interesting to note that in a previous analysis of the $p(\mathrm{e},\mathrm{e}'K^+)\Lambda$ process~\cite{CorthalsHypproc}, based solely on the much smaller dataset released by CLAS in 2003~\cite{Mohring03,Carman03}, we reached identical conclusions concerning the best choice of RPR model. This demonstrates that the RPR approach has considerable predictive power, in spite of its relatively small number of free parameters. 

\begin{figure*}
\begin{center}
\includegraphics*[width=0.85\textwidth]{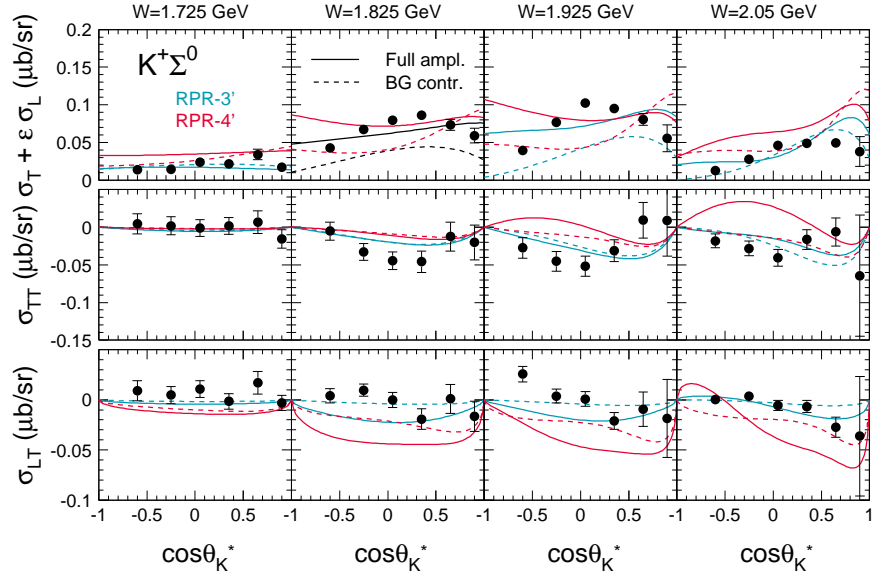}
\end{center}
\caption{$\cos\theta_K^{\ast}$ evolution of the differential cross sections $\sigma_T + \epsilon\, \sigma_L$, $\sigma_{TT}$ and $\sigma_{LT}$ for the $K^+\Sigma^0$ final state at $Q^2 = 0.65$ GeV$^2$, using the two RPR model variants from Table~\ref{tab: kl_models} (full lines) and their respective background contributions (dashed lines). The data are from~\cite{Carman06}.}
\label{fig: k+_sigma_s-new}
\end{figure*}

We have also performed calculations for $p(\mathrm{e},\mathrm{e}'K^+)\Sigma^0$ using the RPR-3$'$ and RPR-4$'$ model variants from Table~\ref{tab: kl_models}. Neither $\sigma_T + \epsilon \, \sigma_L$ nor its separated components were found to exhibit a clear preference for either parameterizarion. The situation is different for the newly measured observables $\sigma_{TT}$ and $\sigma_{LT}$, displayed in Fig.~\ref{fig: k+_sigma_s-new} along with the unseparated cross section. It is clear that RPR-3$'$ performs significantly better than RPR-4$'$ in reproducing the global characteristics of the data. The quality of agreement is, however, considerably worse than for the $K^+\Lambda$ final state, although the absence of any forward peaking of $\sigma_T + \epsilon \, \sigma_L$ is qualitatively reproduced. In contrast to the $p(\mathrm{e},\mathrm{e}'K^+)\Lambda$ reaction, we find relatively large contributions beyond the background, hinting that the $p(\mathrm{e},\mathrm{e}'K^+)\Sigma^0$ channel is more likely to provide interesting resonance information. \\

Summarizing, we have employed a Regge-plus-resonance (RPR) strategy to obtain a description of the $p(\gamma^{(\ast)},K^+)\Lambda,\Sigma^0$ processes in and above the resonance region. The reaction amplitude was constructed from $K^+(494)$ and $K^{\ast +}(892)$ Regge-trajectory exchanges in the $t$ channel, supplemented with a selection of $s$-channel resonances. Apart from the established PDG states, possible contributions of the (as yet) unobserved $D_{13}(1900)$ and $P_{11}(1900)$ resonances were considered. 

Without readjusting any parameter, we compared the various RPR amplitudes constructed in Refs.~\cite{CorthalsL,CorthalsS} for the photoinduced process with the electroproduction data. The electromagnetic form factors of the various $N^{\ast}$ and $\Delta^{\ast}$ states were computed using the Bonn constituent-quark model~\cite{bonn}. 

In the $K\Lambda$ channel, a rotating phase appears to be the optimum choice for both the $K^+(494)$ and $K^{\ast +}(892)$ trajectories. The preferred sign for the $K^{\ast +}(892)$ tensor coupling, which remained undetermined by the photoproduction study, was found to be the negative one. Only the assumption of a missing $D_{13}(1900)$ could be reconciled with the data, whereas the $P_{11}(1900)$ option could be firmly rejected. 

For $K^+\Sigma^0$, only one of the two RPR model variants from the photoproduction study was found to produce acceptable angular dependences for $\sigma_{TT}$ and $\sigma_{LT}$. The best results were obtained with a rotating $K^+(494)$ and constant $K^{\ast +}(892)$ phase, in combination with a negative sign for the $K^{\ast +}(892)$ tensor coupling. 

In comparing the results of this work with those shown in Ref.~\cite{Carman06}, it was observed that models with a reggeized background lead to a better description of the electroproduction data than the background parameterizations typically used in isobar approaches. Furthermore, it was found that most of the $p(\mathrm{e},\mathrm{e}'K)\Lambda,\Sigma^0$ observables can be qualitatively reproduced using a pure $t$-channel Regge model. We believe that the RPR approach provides a powerful tool for interpreting those observables and kinematical regions where additional $s$-channel contributions are required. As a future project, we deem that a combined fit to the $KY$ photo- and electroproduction databases would be useful to further fine-tune the RPR amplitudes. 

\section*{Acknowledgments} 
This work was supported by the Fund for Scientific Research, Flanders (FWO) and the research council of Ghent University.

\clearpage

\begin{table}
\begin{tabular*}{\columnwidth}{@{\extracolsep{\fill}} l l l c c c r}
\hline
& \textbf{RPR} & \textbf{BG model} & $D_{13}$ & $P_{11}$ & $\chi^2_{\mathrm{BG}}$ & \hspace*{-5mm}$\chi^2_{\mathrm{RPR}}$ \\ 
\hline\hline
\underline{$K^+\Lambda$} & RPR-2 & rot.$K$, rot.$K^{\ast}$, $G^t_{K^{\ast}} < 0$ & -- & $\bigstar$ & 16.6 & 3.2 \\
& & & $\bigstar$ & -- & 16.6 & 2.7 \\
& RPR-3 & rot.$K$, cst.$K^{\ast}$, $G^t_{K^{\ast}} > 0$ & -- & $\bigstar$ & 21.7 & 3.1 \\
& & & $\bigstar$ & -- & 21.7 & 3.2 \\
& RPR-4 & rot.$K$, cst.$K^{\ast}$, $G^t_{K^{\ast}} < 0$ & -- & $\bigstar$ & 31.7 & 3.1 \\
& & & $\bigstar$ & -- & 31.7 & 3.1\vspace*{3mm}\\
\underline{$K^+\Sigma^0$} & RPR-3$'$ & rot.$K$, cst.$K^{\ast}$, $G^t_{K^{\ast}} > 0$  & -- & -- & 34.6 & 2.0 \\ 
& RPR-4$'$ & rot.$K$, cst.$K^{\ast}$, $G^t_{K^{\ast}} < 0$  & -- & -- & 8.6 & 2.0 \\
\hline
\end{tabular*}
\caption{RPR variants providing the best description of the $p(\gamma,K^+)\Lambda$ and $p(\gamma,K^+)\Sigma^0$ data from Refs.~\cite{Boyarski69,Vo72,Qui79,McNabb04,Bradford06,Zegers03,Sumihama05,Graal06,Lawall05}. ``Rot.'' and ''cst.'' refer to the rotating or constant Regge trajectory phase. The quoted values of $\chi^2_{\mathrm{RPR}}$ ($\chi^2_{\mathrm{BG}}$) result from a comparison of the full RPR amplitude (Regge background amplitude) to the mentioned set of high-energy and resonance-region data.
}
\label{tab: kl_models}
\end{table} 

\end{document}